\documentstyle[preprint,eqsecnum,aps]{revtex}
\tightenlines
\input epsf

\begin{document}
\draft

\title{The CP violating phase $\delta_{13}$ and the quark mixing angles $\theta_{13}$, $\theta_{23}$ and $\theta_{12}$ from flavor permutational symmetry breaking}

\author{A. Mondrag\'on
   and E. Rodr\'{\i}guez-J\'auregui}
\address{
  Instituto de F\'{\i}sica, UNAM, Apdo. Postal 20-364, 01000 M\'exico,
  D.F. M\'exico.}

\date{\today}
\maketitle
\begin{abstract}
The phase equivalence of the theoretical mixing matrix ${\bf V}^{th}$ derived from the breaking of the flavour permutational symmetry and the standard parametrization ${\bf V}^{PDG}$ advocated by the Particle Data Group is explicitly exhibited. From here, we derive exact explicit expressions for the three mixing angles $\theta_{12}$, $\theta_{13}$, $\theta_{23}$, and the CP violating phase $\delta_{13}$  in terms of the quark mass ratios $(m_{u}/m_{t}, m_{c}/m_{t}, m_{d}/m_{b}, m_{s}/m_{b})$ and the parameters $Z^{*1/2}$ and $\Phi^*$ characterizing the preferred symmetry breaking pattern.
The computed values for the CP violating phase and the mixing angles are: $\delta^*_{13}=75^\circ$, $\sin\theta^*_{12}=0.221$, $\sin\theta^*_{13}=0.0034$, and $\sin\theta^*_{23}=0.040$, which coincide almost exactly with the central values of the experimentally determined quantities.
\end{abstract}

\narrowtext
\section{Introduction}\label{sec:0}
In this paper we are concerned with the functional relations between flavor mixing angles $\theta_{12}$, $\theta_{13}$, $\theta_{23}$, the CP violating phase $\delta_{13}$  and the quark masses resulting from the breaking of the flavor permutational symmetry.\\ 
 In a previous paper ~\cite{ref:1} different Hermitian mass matrices ${\bf M}_q$ of the same modified Fritzsch type were derived from the breaking of the flavor permutational symmetry according to the symmetry breaking scheme 
$S_L(3)\otimes S_R(3)\supset S_L(2)\otimes S_R(2)\supset S_{diag}(2)$. In a symmetry adapted basis, different patterns for the breaking of the permutational symmetry give rise to different mass matrices which differ in the ratio $Z^{1/2}=M_{23}/M_{22}$, and are labeled in terms of the irreducible representations of an auxiliary $\tilde S(2)$ group.
Then, diagonalizing the mass matrices, we obtain exact, explicit expressions for the elements of the mixing matrix ${\bf V^{th}}$, the Jarlskog invariant $J$ and the three inner angles $\alpha$, $\beta$ and $\gamma$ of the unitarity triangle in terms of the quark mass ratios, the symmetry breaking parameter $Z^{1/2}$ and one CP violating phase $\Phi$. The numerical values of $Z^{1/2}$ and $\Phi$ which characterize the preferred symmetry breaking pattern were extracted from a $\chi^2$ fit of the theoretical expressions $|{\bf V}^{th}|$ to the experimentally determined values of the moduli of the elements of the mixing matrix $|{\bf V}^{exp}|$. In this way, we obtained an explicit parametrization of the quark mixing matrix in terms of four quark mass ratios $m_u/m_t$, $m_c/m_t$, $m_d/m_b$, $m_s/m_b$, and the parameters $Z^{1/2}$ and $\Phi$ in excellent agreement with the experimental information about quark mixings and CP violation in the $ K^{\circ}-\bar K^{\circ}$ system and the most recent data on oscillations in the $ B^{\circ}-\bar B^{\circ}$ system. These same experimental data are usually represented by means of the standard parametrization of the mixing matrix \cite{ref:2}, ${\bf V}^{PDG}$, recommended by the Particle Data Group ~\cite{ref:3}, which is written in terms of three mixing angles $\theta_{12}, \theta_{13}, \theta_{23}$ and one CP violating phase $\delta_{13}$. 
The standard parametrization ${\bf V}^{PDG}$, was introduced without taking the functional relations between the quark masses and the flavour mixing parameters into account. In contrast, these functional relations are exactly and explicitly exhibited in the theoretical expressions for ${\bf V^{th}}$ derived in our previous work ~\cite{ref:1}. When the best set of parameters of each parametrization is used, the moduli of corresponding entries of the two parametrizations are numerically equal and give an equally good representation of the experimentally determined values of the moduli of the mixing matrix  
$|V^{exp}_{ij}|$. Hence, we are justified in writing

\begin{equation}\label{eq:1}
|V^{th.}_{ij}|= |V^{PDG}_{ij}|,  
\end{equation}
even though ${\bf V}^{th}$ has only two free, real linearly independent parameters while the number of adjustable parameters in ${\bf V}^{PDG}$ is four.\\

The invariant measurables of the quark mixing matrix are the moduli of its elements, i.e., the quantities $|V_{ij}|$, and the Jarlskog invariant $J$. But even $J$, up to a sign, is a function of the moduli ~\cite{ref:4}. Hence, two different parametrizations, such as $V^{th}_{ij}$ and $V^{PDG}_{ij}$, are equivalent if the moduli of corresponding entries are equal even if the arguments of corresponding entries are different.  This difference is of no physical consequence, it reflects the freedom in choosing the unobservable phases of the quark fields.\\

In this paper, it is shown that a suitable rephasing of the quark fields changes ${\bf V}^{th}$ into a new, phase transformed $\tilde{\bf V}^{th}$ such that all the matrix elements  $\tilde V^{th}_{ij}$ are numerically equal to  the corresponding $V^{PDG}_{ij}$, both in modulus and phase. Once this equality is established, we solve the equations of transformation for  $\sin\theta_{12}$, $\sin\theta_{23}$ and $\sin\theta_{13}$ in terms of the moduli $|\tilde V^{th}_{ij}|$. We also derive exact explicit expressions for the phases of the matrix elements ${V}^{PDG}_{ij}$ in terms of the phases of the matrix elements of  ${V}^{th}_{ij}$. In this way, we derive exact explicit analytical expressions for the mixing parameters $\sin\theta_{12}$, $\sin\theta_{23}$, $\sin\theta_{13}$ and the CP violating phase $\delta_{13}$ of the standard parametrization of the mixing matrix ~\cite{ref:2} in terms of the quark mass ratios $m_u/m_t$, $m_c/m_t$, $m_d/m_b$, $m_s/m_b$, the flavour symmetry breaking parameter $Z^{*1/2}$ and the CP violating phase $\Phi^*$.\\

 The plan of this paper is as follows: In Sec. \ref{sec:2}, we introduce some basic concepts and fix the notation by way of a very brief sketch of the group theoretical derivation of mass matrices with a modified Fritzsch texture. Sect.~\ref{sec:3} is devoted to the derivation of exact, explicit expressions for the elements of the mixing matrix $V^{th}_{ij}$ in terms of the quark mass ratios and the parameters $Z^{1/2}$ and  $\Phi$ characterizing the symmetry breaking pattern. In Sec.~\ref{sec:4}, the phase equivalence of ${\bf V}^{th}$ and ${\bf V}^{PDG}$ is explicitly exhibited, and a set of equations expressing the non-vanishing arguments $w^{PDG}_{ij}$ of $V^{PDG}_{ij}$ in terms of the arguments $w^{th}_{ij}$ of $V^{th}_{ij}$ is derived.  
Explicit expressions for the mixing parameters $\sin\theta_{12}$, $\sin\theta_{23}$, $\sin\theta_{13}$ and the CP violating phase $\delta_{13}$ as functions of the quark mass ratios and the parameters  $Z^{*1/2}$ and $\Phi^*$ characterizing the preferred symmetry breaking scheme are obtained in Sec.~\ref{sec:5} and ~\ref{sec:6}. Our paper ends in Sec.~\ref{sec:7} with a summary of results and some conclusions.

\section{Mass matrices from the breaking of $S_{L}(3)\otimes S_{R}(3)$ }
\label{sec:2}

In the Standard Model, analogous fermions in different generations,
say ${\it u,c}$ and ${\it t}$ or ${\it d,s}$ and ${\it b}$, have
completely identical couplings to all gauge bosons of the strong, weak
and electromagnetic interactions.  Prior to the introduction of the
Higgs boson and mass terms, the Lagrangian is chiral and invariant
with respect to any permutation of the left and right quark fields.
The introduction of a Higgs boson and the Yukawa couplings give mass
to the quarks and leptons when the gauge symmetry is spontaneously
broken. The quark mass term in the Lagrangian, obtained by taking the
vacuum expectation value of the Higgs field in the quark Higgs
coupling, gives rise to quark mass matrices ${\bf M_d}$ and ${\bf
  M_u}$,

\begin{equation}\label{eq:21}
{\cal L}_{Y} ={\bf \bar{q}}_{d,L}{\bf M}_{d}{\bf q}_{d,R}+
{\bf\bar{q}}_{u,L}{\bf M}_{u}{\bf q}_{u,R}+h.c.
\end{equation}

In this expression, ${\bf q}_{d,L,R}(x)$ and ${\bf q}_{u,L,R}(x)$ denote the
left and right quark $d$- and $u$-fields in the current or weak 
basis, ${\bf q}_{q}(x)$ is a column matrix, its components ${\bf q}_{q,k}(x)$ are the quark Dirac fields, $k$ is the flavour index. In this basis, the charged hadronic currents,

\begin{equation}
\label{eq:23}
J_{\mu}\sim \bar{q}_{u,L}\gamma _{\mu}q_{d,L},
\end{equation}
are not changed if both, the d-type and the u-type fields are transformed with the same unitary matrix.

\subsection{Modified Fritzsch texture}
\label{sec:2.1}
A number of authors \cite{ref:1}, [\cite{ref:5}-\cite{ref:23}] have pointed out that realistic quark mass matrices result from the flavour permutational symmetry $S_{L}(3)\otimes S_{R}(3)$ and its spontaneous or explicit breaking.
The group $S(3)$ treats three objects symmetrically, while the
hierarchical nature of the mass matrices is a consequence of the
representation structure $\bf{1\oplus2}$ of $S(3)$, which treats the
generations differently.
Under exact $S_{L}(3)\otimes S_{R}(3)$ symmetry, the mass spectrum for either up or down quark sectors consists of one massive particle in a singlet irreducible representation and a pair of massless particles in a doublet irreducible representation, the corresponding quark mass matrix with the exact $S_{L}(3)\otimes S_{R}(3)$ symmetry will be denoted by ${\bf M}_{3q}$. In order to generate masses for the first and second families, we add the terms ${\bf M}_{2q}$ and ${\bf M}_{1q}$ to ${\bf M}_{3q}$. The term ${\bf M}_{2q}$ breaks the permutational symmetry  $S_{L}(3)\otimes S_{R}(3)$ down to $S_{L}(2)\otimes S_{R}(2)$ and mixes the singlet and doublet representation of $S(3)$. ${\bf M}_{1q}$ transforms as the mixed symmetry term in the doublet complex tensorial representation of $S_{diag}(3) \subset S_{L}(3)\otimes S_{R}(3)$. Putting the first family in a complex representation will allow us to have a CP violating phase in the mixing matrix. Then, in a symmetry adapted basis , ${\bf M}_{q}$ takes the form

\begin{equation}\label{eq:2.11}
\begin{array}{rcl}
{M_{q}}&=&{m_{3q}}\left[ \pmatrix{
0 & {A_{q}}e^{-i\phi_{q}} & 0 \cr
{A_{q}}e^{i\phi_{q}} & 0 & 0 \cr
0 & 0 & 0 \cr
}+\pmatrix{
0 & 0 & 0 \cr
0 & -\triangle_{q}+\delta_{q} & B_{q} \cr
0 & B_{q} & \triangle_{q}-\delta_{q} \cr
}\right]\\ &+&m_{3q}\pmatrix{
0 & 0 & 0 \cr
0 & 0 & 0 \cr
0 & 0 & 1-\triangle_{q} \cr
}=m_{3q}\pmatrix{
0 & A_{q}e^{-i\phi_{q}} & 0 \cr
A_{q}e^{i\phi_{q}} & -\triangle_{q}+\delta_{q} & B_{q} \cr
0 & B_{q} & 1-\delta_{q} \cr
}.
\end{array}
\end{equation}

From the strong hierarchy in the masses of the quark families, 
$m_{3q}>> m_{2q}> m_{1q}$, we expect $1-\delta_{q}$ to be very close to unity.
The entries in the mass matrix may be readily expressed in terms of the mass eigenvalues $(m_{1q}, -m_{2q}, m_{3q})$ and the small parameter $\delta_{q}$.
 Computing the invariants of $M_{q}$, $tr M_{q}$, $tr {M_{q}}^{2}$ and $det M_{q}$, we get
\begin{eqnarray}\label{eq:2.12}
A^{2}_{q}={\tilde m_{1q}}{\tilde m_{2q}}(1-\delta _{q})^{-1}\qquad ,\qquad
\triangle _{q}= {\tilde m_{2q}}-{\tilde m_{1q}},\\ \cr
B^{2}_{q}=\delta_{q}((1-\tilde m_{1q}+\tilde m_{2q}-\delta _{q})-
\tilde m_{1q}{\tilde m_{2q}}(1-\delta _{q})^{-1}),
\end{eqnarray}
where 
${\tilde m_{1q}}={m_{1q}/m_{3q}}$ and 
${\tilde m_{2q}}={m_{2q}/m_{3q}}$.

If each possible symmetry breaking pattern is now characterized by the ratio
\begin{equation}\label{eq:2.13}
{Z_{q}}^{1/2}={B_{q}/(-\triangle _{q}+\delta_{q})},
\end{equation}
the small parameter $\delta _{q}$ is obtained as the solution of the cubic equation

\begin{equation}\label{eq:2.14}
\delta_{q}\left[ (1+\tilde m_{2q}- \tilde m_{1q}- \delta_{q})(1-\delta_{q})-
{\tilde m_{1q}}{\tilde m_{2q}}\right] - Z_{q}(-{\tilde m_{2q}}+{\tilde m_{1q}}+
\delta_{q})^{2}=0,
\end{equation}
which vanishes when $Z_{q}$ vanishes. An exact explicit expression for $\delta_q$ as function of the quark mass ratios and $Z_q$ is given in \cite{ref:1}. 
 An approximate solution to Eq. (\ref{eq:2.14}) for $\delta_{q}(Z_{q})$, valid for small
values of $Z_{q}$ ($Z_{q}\leq 10$), is
\begin{eqnarray}\label{eq:2.15}
 \delta_{q}\left( Z_{q} \right)\approx {Z_{q}
\left(   \tilde{m}_{2q}-\tilde{m}_{1q} \right)^{2}
\over \left(1-\tilde{m}_{1q} \right)\left( 1  +
\tilde{m}_{2q} \right)+2Z_{q}\left(   \tilde{m}_{2q}-\tilde{m}_{1q} 
\right)(1+{1\over 2}(\tilde{m}_{2q}-\tilde{m}_{1q}))}~.
\end{eqnarray}
\subsection{Symmetry breaking pattern}
\label{sec:2.2}
In the symmetry adapted basis, the matrix $\bf{M}_{2q}$, written in term of $Z^{1/2}_q$, takes the form

\begin{equation}\label{eq:2.21}
{\bf M}_{2q} =
m_{3q}\left(-\tilde{m}_{2q}+\tilde{m}_{1q}+\delta_{q}\right)\pmatrix{
0 & 0 & 0 \cr
0 & 1 & Z^{1/2}_q \cr
0 & Z^{1/2}_q & -1 \cr
},
\end{equation}
when $Z^{1/2}_q$ vanishes, ${\bf M}_{2q}$ is diagonal and there is no mixing of singlet and doublet representations of $S(3)$. Therefore, in the symmetry adapted basis, the parameter $Z^{1/2}_q$ is a measure of the amount of mixing of singlet and doublet irreducible representations of $S_{diag}(3)\subset S_L(3) \otimes S_R(3)$.\\

We may easily give a meaning to $Z^{1/2}_q$ in terms of permutations. From Eqs. (\ref{eq:21}) and (\ref{eq:2.21}), we notice that the symmetry breaking term in the Yukawa Lagrangian, $\bar{\bf q}_L {\bf M}_{2q}{\bf q}_R $ is a functional of only two fields:
$\frac{1}{\sqrt 3}\left( q_2(X)+\sqrt{2} q_3(X)\right)$ and $\frac{1}{\sqrt 3}\left(-\sqrt{2} q_2(X)+ q_3(X)\right)$.
Under the permutation of these fields, $\bar{\bf q}_L {\bf M}_{2q}{\bf q}_R $ splits into the sum of an antisymmetric term $\bar{\bf q}_L {\bf M}^A_{2q}{\bf q}_R $ which changes sign, and a symmetric term $\bar{\bf q}_L {\bf M}^S_{2q}{\bf q}_R $, which remains invariant,
\begin{equation}\label{eq:2.23}
{\bf M}_{2q}=-\frac{2}{9}m_{3q}\bigg\{a\pmatrix{
0 & 0 & 0 \cr
0 & 1 & -\sqrt 8 \cr
0 & -\sqrt 8 & -1 \cr
}+ 2b\pmatrix{
0 & 0 & 0 \cr
0 & 1 & \frac{1}{\sqrt 8} \cr
0 & \frac{1}{\sqrt 8} & -1 \cr
}\bigg\},
\end{equation}
where $a=(\delta_q-\triangle_q)(\sqrt 2 Z^{1/2}_q -\frac{1}{2})$ and $b=(\delta_q-\triangle_q)(\frac{\sqrt 2}{2} Z^{1/2}_q +{2})$.
It is evident that there is a corresponding decomposition of the mixing parameter $Z^{1/2}_q$,
\begin{equation}\label{eq:2.25}
{Z_{q}}^{1/2}=N_{Aq}Z^{1/2}_A+N_{Sq}Z^{1/2}_S
\end{equation}
 with 

\begin{equation}\label{eq:2.27}
1=N_{Aq}+N_{Sq},
\end{equation}
where $Z^{1/2}_A=-\sqrt 8$ is the mixing parameter of the matrix ${\bf M}^A_{2q}$, and $Z^{1/2}_S=\frac{1}{\sqrt 8}$ is the mixing parameter of ${\bf M}^S_{2q}$.
In this way, a unique linear combination of $Z^{1/2}_{A}$ and $Z^{1/2}_{S}$  is associated to the simmetry breaking pattern characterized by $Z^{1/2}_{q}$. Thus, the different symmetry breaking patterns defined by ${\bf M}_{2q}$ for different values of the mixing parameter $Z^{1/2}_{q}$ are labeled in terms of the irreducible representations of the group $\tilde S(2)$ of permutations of the two fields in $\bar{\bf q}_L {\bf M}_{2q}{\bf q}_R $. 
The pair of numbers $(N_A, N_S)$ enters as a convenient mathematical label of the symmetry breaking pattern without introducing any assumption about the actual pattern of $S_L(3)\otimes S_R(3)$ symmetry breaking realized in nature.
\subsection{The Jarlskog invariant}
\label{sec:2.3}

The Jarlskog invariant, $J$, may be computed directly from the commutator of the mass matrices ~\cite{ref:4}

\begin{equation}\label{eq:2.31}
J=- \frac{det \{ -i[{\bf M}_{u}, {\bf M}_{d}]\} }{F}
\end{equation}
where 
\begin{equation}\label{eq:2.33}
F=(1+\tilde{m}_{c})(1-\tilde{m}_{u})(\tilde{m}_{c}+\tilde{m}_{u})
(1+\tilde{m}_{s})(1-\tilde{m}_{d})(\tilde{m}_{s}+\tilde{m}_{d}).
\end{equation}
Substitution of the expression (\ref{eq:2.11}) for ${\bf M}_u$ and ${\bf M}_d$, in Eq. (\ref{eq:2.31}), with $Z^{1/2}_u=Z^{1/2}_d=Z^{1/2}$ gives

\begin{eqnarray}\label{eq:2.35}
&J&={{Z  \sqrt{{\tilde m_{u}/ \tilde m_{c}}
\over{1-\delta_{u}}} \sqrt{{\tilde m_{d}/ \tilde m_{s}}
\over {1-\delta_{d}}}sin{\Phi}}\over{(1+\tilde m_{c})
(1-\tilde m_{u})(1+\tilde m_{u}/ \tilde m_{c})
(1+\tilde m_{s})(1-\tilde m_{d})(1+\tilde m_{d}/ \tilde m_{s})}}\cr
& \times &\bigg\{
 [(-\triangle_{u}+\delta_{u})(1-\delta_{d})-(-\triangle_{d}+\delta_{d})(1-\delta_{u})]^{2}-{\left({\tilde m_{u}\tilde m_{c}}\over {1-\delta_{u}}\right)}(-\triangle_{d}+\delta_{d})^{2}\cr & - & {\left({\tilde m_{d}\tilde m_{s}\over {1-\delta_{d}}}\right)}(-\triangle_{u}+\delta_{u})^{2}+2 \sqrt{\tilde m_{u}\tilde m_{c}\over {1-\delta_{u}}}\sqrt{\tilde m_{d}\tilde m_{s}\over {1-\delta_{d}}}(-\triangle_{u}+\delta_{u})(-\triangle_{d}+\delta_{d})cos{\Phi}\bigg\}.
\end{eqnarray}
where $\triangle _q$ and $\delta_q$ are defined in Eqs. (\ref{eq:2.12}) and (\ref{eq:2.14}).
In this, way, an exact closed expression for $J$ in terms of the quark mass ratios, the CP violating phase $\Phi$, and the parameter $Z$ that characterizes the symmetry breaking pattern is derived.

\section{The Mixing Matrix}\label{sec:3}
The Hermitian mass matrix ${\bf M}_{q}$ may be written in terms of a real symmetric matrix ${\bf\bar M}_{q}$ and a diagonal matrix of phases ${\bf P}_{q}$ as follows

\begin{equation}\label{eq:31}
{\bf M}_{q}={\bf P}_{q}{{\bf\bar M}_{q}}{{\bf P}_{q}}^{\dagger},
\end{equation}
The real symmetric matrix ${\bf\bar M}_{q}$ may be brought to a diagonal form by means of an orthogonal transformation 
\begin{equation}\label{eq:33}
{\bf\bar M}_{q}={\bf O}_{q}{\bf M}_{q, diag}{\bf O}^{T}_{q},
\end{equation}
where
\begin{equation}\label{eq:35}
{\bf M}_{q, diag}=m_{3q}~diag\left[ ~\tilde m_{1q},~-\tilde m_{2q}, ~1\right],
\end{equation}
with subscripts $1, 2, 3$ refering to $u, c, t$ in the u-type sector and $d, s , b$ in the d-type sector.\\
After diagonalization of the mass matrices ${\bf M}_q$, one obtains the mixing matrix ${\bf V}^{th}$ as

\begin{equation}\label{eq:37}
{\bf V}^{th}={{\bf O}_{u}}^{T}{\bf P}^{u-d}{\bf O}_{d}, 
\end{equation}
where ${\bf P}^{u-d}$ is the diagonal matrix of relative phases

\begin{equation}\label{eq:39}
{\bf P}^{u-d}=diag[1,e^{i\Phi},e^{i\Phi}],
\end{equation} 
and 
\begin{equation}\label{eq:311}
\Phi=(\phi_{u}-\phi_{d}).
\end{equation} 
The orthogonal matrix ${\bf O}_{q}$is given by

\begin{equation}\label{eq:313}
\begin{array}{rcl}
{\bf O}_{q}= \pmatrix{
({\tilde m_{2q}}f_{1}/D_{1})^{1/2} & - ({\tilde m_{1q}}f_{2}/D_{2})^{1/2} & 
({\tilde m_{1q}}{\tilde m_{2q}}f_{3}/D_{3})^{1/2} \cr 
((1-\delta_{q}){\tilde m_{1q}}f_{1}/D_{1})^{1/2} & ((1-\delta_{q}){\tilde m_{2q}}f_{2}/D_{2})^{1/2} & ( (1-\delta_{q}) f_{3}/D_{3})^{1/2} \cr
-({\tilde m_{1q}}f_{2}f_{3}/D_{1})^{1/2} & 
-({\tilde m_{2q}}f_{1}f_{3}/D_{2})^{1/2} & ( f_{1}f_{2}/D_{3})^{1/2} \cr 
},
\end{array}
\end{equation}
where
\begin{eqnarray}\label{eq:315}
{\rm {f}}_{1}=1-\tilde{m}_{1q}-{\delta}_{q},\quad\quad
{\rm {f}}_{2}=1+\tilde{m}_{2q}-{\delta}_{q},\quad\quad
{\rm {f}}_{3}={\delta}_{q},
\end{eqnarray}

\begin{eqnarray}\label{eq:317}
{D}_{1}=(1-\delta_q)\left( 1-\tilde{m}_{1q} \right)\left( 
\tilde{m}_{2q}+\tilde{m}_{1q} \right),
\end{eqnarray}

\begin{eqnarray}\label{eq:319}
{D}_{2}=(1-\delta_q)\left( 1+\tilde{m}_{2q} \right)\left( 
\tilde{m}_{2q}+\tilde{m}_{1q} \right),
\end{eqnarray}

\begin{eqnarray}\label{eq:321}
{D}_{3}=(1-\delta_q)\left( 1+\tilde{m}_{2q} \right)\left( 
1-\tilde{m}_{1q} \right).
\end{eqnarray}

In these expressions, $\delta_u$ and $\delta_d$ are, in principle, functions of the quark mass ratios and the parameters $Z^{1/2}_u$ and $Z^{1/2}_d$ respectively. However, in \cite{ref:1} we found that keeping  $Z^{1/2}_u$ and $Z^{1/2}_d$ as free, independent parameters gives rise to a continuous ambiguity in the fitting of $|V^{th}_{ij}|$ to the experimental data. To avoid this ambiguity we further assumed that the up and down mass matrices are generated following the same symmetry breaking pattern, that is,
\begin{equation}\label{eq:323}
Z^{1/2}_u=Z^{1/2}_d=Z^{1/2}.
\end{equation}
Then, from Eqs. (\ref{eq:37}) - (\ref{eq:323}) all matrix elements in ${\bf V}^{th}$ may be written in terms of four quark mass ratios and only two free, real parameters: the parameter $Z^{1/2}$ which characterizes the symmetry breaking pattern in the u- and d-sectors and the CP violating phase $\Phi$.
The computation of $V^{th}_{ij}$ is quite straightforward. Here, we will give, in explicit form, only those elements of  ${\bf V}^{th}$ which will be of use later. From Eqs. (\ref{eq:37})-(\ref{eq:323}) we obtain, 
\begin{eqnarray}\label{eq:325}\hspace{1.5cm}
V^{th}_{us}&=&-\left(\frac{ \tilde {m}_c \left(1-\tilde{m}_u -\delta_u \right)
\tilde {m}_d\left(1+\tilde{m}_s -\delta_d \right) }
{\left( 1-\delta_u \right) \left( 1-\tilde{m}_u \right)
\left( \tilde{m}_c +\tilde{m}_u \right)\left(1-\delta_d \right)
\left( 1+\tilde{m}_s\right)
\left( \tilde{m}_s + \tilde{m}_d\right)}\right)^{1/2} \cr &+&
\left( \frac{\tilde{m}_u \tilde{m}_s}
{\left( 1-\tilde{m}_u \right)\left( \tilde{m}_c+
\tilde{m}_u \right)\left(\tilde{m}_d +\tilde{m}_s
\right)} \right) ^{1/2}
\bigg\{ \left(\frac {\left(1-\tilde{m}_u - \delta_u \right)
\left(1+\tilde{m}_s- \delta_d \right)}{\left( 1+\tilde{m}_s \right)}
\right)^{1/2}\cr &+&
\left(\frac{ \left(1+\tilde{m}_c- \delta_u \right)\delta_u
\left(1-\tilde{m}_d - \delta_d \right)\delta_d }
{(1-\delta_u)(1-\delta_d)(1+\tilde{m}_s)}\right)^{1/2}
\bigg\}e^{i\Phi}
\end{eqnarray}

\begin{eqnarray}\label{eq:327}
V^{th}_{ub}&=&\left( \frac{\tilde {m}_c (1-\tilde{m}_u -\delta_u ) }{ (1-\delta_u)
(1-\tilde{m}_u)(\tilde{m}_c + \tilde{m}_u)} \frac {\tilde {m}_d \tilde {m}_s
\delta_d }{(1- \delta_d )(1+ \tilde {m}_s)(1 - \tilde {m}_d) } \right)^{1/2}
\cr
&+&\bigg\{- \left(\frac{\tilde{m}_u (1+\tilde{m}_c -\delta_u)\delta_u 
(1-\tilde{m}_d -\delta_d)(1+\tilde{m}_s -\delta_d)}
{(1- \delta_u )(1-\tilde{m}_u)(\tilde{m}_c +\tilde{m}_u)(1- \delta_d)
(1+\tilde{m}_s)(1-\tilde{m}_d)}
\right)^{1/2}
\cr &+& \left(\frac{\tilde{m}_u (1-\tilde{m}_u -\delta_u)\delta_d}
{(1-\tilde{m}_u)(\tilde{m}_c +\tilde{m}_u)(1+\tilde{m}_s)(1-\tilde{m}_d)}
\right)^{1/2}\bigg\}e^{i\Phi}  
\end{eqnarray}

\begin{eqnarray}\label{eq:329}
V^{th}_{cs}&=&\left(\frac{\tilde {m}_u \left( 1+\tilde{m}_c -\delta_u \right)
\tilde {m}_d \left( 1+\tilde{m}_s -\delta_d \right)}{\left( 1-\delta_u \right)
\left( 1+\tilde{m}_c \right)\left( \tilde {m}_c+\tilde {m}_u \right)
\left( 1-\delta_d \right)\left( 1+\tilde{m}_s \right)
\left( \tilde {m}_s+\tilde {m}_d \right)}\right)^{1/2}\cr &+& \bigg\{
\left(\frac{\tilde {m}_c\delta_u\left( 1-\tilde {m}_u-\delta_u \right)
\tilde {m}_s\delta_d
\left( 1-\tilde{m}_d -\delta_d \right)}{\left( 1-\delta_u \right)
\left( 1+\tilde{m}_c \right)\left( \tilde {m}_c+\tilde {m}_u \right)
\left( 1-\delta_d \right)\left( 1+\tilde{m}_s \right)
\left( \tilde {m}_s+\tilde {m}_d \right)}\right)^{1/2}\cr &+&
\left(\frac{\tilde {m}_c
\left( 1+\tilde{m}_c -\delta_u \right)\tilde {m}_s
\left( 1+\tilde{m}_s -\delta_d \right)}{
\left( 1+\tilde{m}_c \right)\left(1-\tilde {m}_u \right)
\left( 1+\tilde{m}_s \right)
\left( 1-\tilde {m}_d \right)}\right)^{1/2}
\bigg\}e^{i\Phi}
\end{eqnarray}

and 

\begin{eqnarray}\label{eq:331}
{V^{th}_{cb}}&=&- \left( \frac{\tilde {m}_u (1+\tilde{m}_c -\delta_u ) }
{ (1-\delta_u)(1+\tilde{m}_c)(\tilde{m}_c + \tilde{m}_u)} \frac {
\tilde {m}_d \tilde {m}_s \delta_d }{(1- \delta_d )(1+ \tilde {m}_s)
(1 - \tilde {m}_d) } \right)^{1/2}\cr
&+& \bigg\{- \left(\frac{\tilde{m}_c (1-\tilde{m}_u -\delta_u)\delta_u 
(1-\tilde{m}_d -\delta_d)(1+\tilde{m}_s -\delta_d)}
{(1- \delta_u )(1+\tilde{m}_c)(\tilde{m}_c +\tilde{m}_u)(1- \delta_d)
(1+\tilde{m}_s)(1-\tilde{m}_d)}
\right)^{1/2} 
\cr &+&\left(\frac{\tilde{m}_c (1+\tilde{m}_c -\delta_u)}
{(\tilde{m}_c +\tilde{m}_u)(1+\tilde{m}_c)}\frac{\delta_d}{(1+\tilde{m}_s)
(1-\tilde{m}_d)}\right)^{1/2}\bigg\}e^{i\Phi}.
\end{eqnarray}
\subsection{The ``best'' symmetry breaking pattern}
\label{sec:3.1}
In order to find the actual pattern of $S_L(3)\otimes S_R(3)$ symmetry breaking realized in nature,we made a $\chi^2$ fit of the exact expressions for the moduli of the entries in the mixing matrix, $|V^{th}_{ij}|$, the Jarlskog invariant, $J^{th}$, and the three inner angles of the unitarity triangle, $\alpha^{th}$, $\beta^{th}$ and $\gamma^{th}$, to the experimentally determined values of $|{ V}^{exp}_{ij}|$, $J^{exp}$, $\alpha^{exp}$, $\beta^{exp}$ and $\gamma^{exp}$. A detailed account of the fitting procedure is given in \cite{ref:1}. Here, we will give only a brief relation of the main points in the fitting procedure.\\
For the purpose of calculating quark mass ratios and computing the mixing matrix, it is convenient to give all quark masses as running masses at some common energy scale \cite{ref:24}, \cite{ref:25}. In the present calculation, following Peccei \cite{ref:24}, Fritzsch \cite{ref:26} and the Ba-Bar book \cite{ref:27}, we used the values of the running quark masses evaluated at $\mu=m_t$.

\begin{eqnarray}\label{eq:3.10}
m_u&=&3.25\pm 0.9~MeV\quad\quad 
m_c=760\pm 29.5~MeV\quad\quad 
m_t=171.0\pm 12~GeV
\cr
m_d&=&4.4\pm 0.64~MeV\quad\quad~~
m_s=100\pm 6~MeV\quad\quad~~ 
m_b=2.92\pm 0.11~GeV
\end{eqnarray}
These values, with the exception of $m_s$, $m_c$ and $m_b$, were taken from the work of Fusaoka and Koide \cite{ref:25} see also Fritzsch \cite{ref:26} and Leutwyler \cite{ref:28}. The values of $m_c(m_t)$ and $m_b(m_t)$ were obtained by rescaling to $\mu=m_t$ the recent calculations of $m_c(m_c)$ and $m_b(m_b)$ by Pineda and Yndur\'ain \cite{ref:29} and Yndur\'ain \cite{ref:30}. The value of $m_s$
 agrees with the latest determination made by the ALEPH collaboration from a study of $\tau$ decays involving kaons \cite{ref:31}. \\
We kept the mass ratios $\tilde m_c=m_c/m_t$ , $\tilde m_s=m_s/m_b$ and $\tilde m_d=m_d/m_b$ fixed at their central values

\begin{eqnarray}\label{eq:3.11}
\tilde m_c=0.0044, \quad\quad\tilde m_s=0.034 \quad\quad ~and 
\quad\quad\tilde m_d=0.0015,
\end{eqnarray}
but we took the value 

\begin{eqnarray}\label{eq:3.13}
\tilde m_u=0.000032,
\end{eqnarray}
which is close to its upper bound.
We found the following best values for $\Phi$ and $Z^{1/2}$,

\begin{eqnarray}
\Phi^{*}=90^{\circ},\quad\quad Z^{*1/2}=\frac{1}{2}\left[Z^{1/2}_S-Z^{1/2}_A\right]=\sqrt\frac{81}{32}.
\label{eq:3.15}
\end{eqnarray}
corresponding to a value of $\chi^2\leq 0.32$.  The values of the parameters $\delta_u(Z)$ and $\delta_d(Z)$ obtained from (\ref{eq:3.11}), (\ref{eq:3.13}) and  (\ref{eq:3.15}) are
\begin{eqnarray}\label{eq:3.16}
\delta_u(Z^{*1/2})=0.000048, \quad\quad\quad \delta_d(Z^{*1/2})=0.00228.
\end{eqnarray}
Before proceeding to give the numerical results for the mixing matrix ${\bf V}^{th}$, it will be convenient to stress the following points:
\begin{enumerate}
\item
The masses of the lighter quarks are the less well determined, while the moduli of the entries in $|V^{exp}_{ij}|$ with the largest error bars, namely $|V_{ub}|$ and $|V_{td}|$, are the most sensitive to changes in the ratios $m_u/m_c$ and $m_d/m_s$ respectively. Hence, the quality of the fit of $|V^{th}_{ij}|$ to $|V^{exp}_{ij}|$ is good $(\chi^2\leq 0.5)$ even if relatively large changes in the masses of the lighter quarks are made. The sensitivity of $|V_{ub}|$ and $|V_{td}|$ to changes in  $m_u/m_c$ and $m_d/m_s$ respectively, is reflected in the shape of the unitarity triangle which changes appreciably when the masses of the ligther quarks change within their uncertainty ranges. The best simultaneous $\chi^2$ fit of $|V^{th}_{ij}|$, $J^{th}$, and $\alpha^{th}$, $\beta^{th}$ and $\gamma^{th}$, to the experimentally determined quantities was obtained when the ratio  $\tilde m_u=m_u/m_t$ is taken close to its upper bound, as given in (\ref{eq:3.13}). Furthermore, the chosen high value of $\tilde m_u$ gives for the ratio $|V_{ub}/V_{cb}|$ the value 

\begin{eqnarray}\label{eq:3.17}
\frac{|V_{ub}|}{|V_{cb}|}\approx \sqrt{\frac{m_u}{m_c}}=0.085\pm 0.009
\end{eqnarray}
in very good agreement with its latest world average [\cite{ref:32}, \cite{ref:33}, \cite{ref:34}].
\item
As the energy scale changes, say from $\mu=m_t$ to $\mu=1~GeV$, the running quark masses change appreciably, but since the masses of light and heavy quarks increase almost in the same proportion, the resulting dependence of the quark mass ratios on the energy scale is very weak. When the energy scale changes from  $\mu=m_t$ to $\mu=1~GeV$, $\tilde m_u$ and $\tilde m_d$ decrease by about $25\%$ and $\tilde m_c$ and $\tilde m_s$ also decrease but by less than $16\%$.
\item
In view of the previous considerations, a reasonable range of values for the running quark mass ratios, evaluated at $\mu=m_t=171~GeV$, would be as follows
\begin{eqnarray}\label{eq:3.19}
0.000022\leq &\tilde m_u&\leq 0.000037\cr
0.0043\leq &\tilde m_c&\leq 0.0046\cr
0.0013\leq &\tilde m_d&\leq 0.0017\cr
0.032\leq &\tilde m_s&\leq 0.036
\end{eqnarray}
\end{enumerate}
  
The results of the $\chi^2$ fit of the theoretical expressions for $|{V}^{th}_{ij}|$, $J^{th}$, $\alpha^{th}$, $\beta^{th}$ and $\gamma^{th}$ to the experimentally determined quantities is as follows:\\
The quark mixing matrix computed from the theoretical expresion ${\bf V}^{th}$ with the numerical values of quark mass ratios given in (\ref{eq:3.11})and (\ref{eq:3.13}) and the corresponding best values of the symmetry breaking parameter, $Z^{*1/2}=\sqrt{81/32}$, and the CP-violating phase, $\Phi^*=90^{\circ}$, is 
 
\begin{equation}
V^{th}= \pmatrix{
0.9753~e^{i1^{\circ}} & 0.221~e^{i158^{\circ}} & 0.0034~e^{i84^{\circ}}\cr
0.220~e^{i112^{\circ}} & 0.9745~e^{i89^{\circ}} & 0.040~e^{i90^{\circ}} \cr
0.0085~e^{i270^{\circ}} & 0.039~e^{i270^{\circ}} & 0.9992~e^{i90^{\circ}} \cr
}
\label{eq:3.21}
\end{equation}
In order to have an estimation of the sensivity of our numerical results to the uncertainty in the values of the quark mass ratios, we computed the range of values of the matrix of moduli $|V^{th}_{ij}|$, corresponding to the range of values of the mass ratios given in (\ref{eq:3.19}), but keeping $\Phi$ and $Z^{1/2}$ fixed at the values $\Phi^*=90^{\circ}$ and $Z^{*1/2}=\sqrt{81/32}$.
The result is
\begin{equation}
|{\bf V}^{th}|=\pmatrix{
0.9735 - 0.9771 & 0.2151 - 0.2263 & 0.0028 - 0.0040 \cr
0.2151 - 0.2263 & 0.9726 - 0.9764 & 0.037 - 0.043 \cr
0.0078 - 0.0093 & 0.036 - 0.042 & 0.9991 - 0.9993 \cr
},
\label{eq:3.23}
\end{equation}
which is to be compared with the experimentally determined values of the matrix of moduli \cite{ref:3},
\begin{equation}
|{\bf V}^{exp}|=\pmatrix{
0.9745 - 0.9760 & 0.217 - 0.224 & 0.0018 - 0.0045 \cr
0.217 - 0.224 & 0.9737 - 0.9753 & 0.036 - 0.042 \cr
0.004 - 0.013 & 0.035 - 0.042 & 0.9991 - 0.9994 \cr
}.
\label{eq:3.25}
\end{equation}

As is apparent from (\ref{eq:3.21}), (\ref{eq:3.23}) and (\ref{eq:3.25}), the agreement between computed and experimental values of all entries in the mixing matrix is very good. The estimated range of variation in the computed values of the moduli of the four entries in the upper left corner of the matrix  $|{\bf V}^{th}|$ is larger than the error band in the corresponding entries of the matrix of the experimentally determined values of the moduli $|{\bf V}^{exp}|$.  The estimated range of variation in the computed values of the entries in the third column and the third row of $|{V}^{th}_{ij}|$ is comparable with the error band of the corresponding entries in the matrix of experimentally determined values of the moduli, with the exception of the elements $|V^{th}_{ub}|$ and $|V^{th}_{td}|$ in which case the estimated range of variation due to the uncertainty in the values of the quark mass ratios is significantly smaller than the error band in the experimentally determined value of $|V^{exp}_{ub}|$ and $|V^{exp}_{td}|$.\\

The value obtained for the Jarlskog invariant is 
\begin{equation}\label{eq:3.29}
J^{th}=2.8\times10^{-5}
\end{equation}
in good agreement with the value $|J^{exp}|=(3.0\pm1.3)\times 10^{-5}sin\delta$ obtained from current data on CP violation in the $K^{\circ}-\bar K^{\circ}$ mixing system \cite{ref:3} and the $B^{\circ}-\bar B^{\circ}$ mixing system \cite{ref:27}.\\

For the inner angles of the unitarity triangle, we found the following central values:
\begin{eqnarray}\label{eq:3.27}
\alpha=83^{\circ}\quad\quad\beta=22^{\circ}\quad\quad\gamma=75^{\circ}.
\end{eqnarray}
An estimation of the range of values of the three inner angles of the unitarity triangle, compatible with the experimental information on the absolute values of the matrix elements ${\bf V}^{exp}$, is given by Mele \cite{ref:32} and Ali \cite{ref:33}. According to this authors, $79^{\circ}\leq\alpha\leq 102^{\circ}$, $21^{\circ}\leq\beta\leq 28^{\circ}$, and $55^{\circ}\leq\gamma\leq 78^{\circ}$. We see that the central value of $\beta$ obtained in this work is close to the lower limit according to Mele \cite{ref:32}, while $\gamma$ is close to the upper limit given by Mele \cite{ref:32} and $\alpha$ is in the allowed range given by these authors.

\section{Phase equivalence of ${\bf V}^{th}$ and the standard parametrization ${\bf V}^{PDG}$} 
\label{sec:4}

The standard parametrization ~\cite{ref:2} of the mixing matrix recomended by the Particle Data Group \cite{ref:3} is written in terms of three mixing angles $\theta_{12}, \theta_{23}, \theta_{13}$ and one CP violating phase $\delta_{13}$, 

\begin{equation}\label{eq:4431}
{\bf V}^{PDG}=\pmatrix{
c_{12}c_{13} & s_{12}c_{13} & s_{13}e^{-i\delta_{13}} \cr
-s_{12}c_{23}-c_{12}s_{23}s_{13}e^{i\delta_{13}} & c_{12}c_{23}-s_{12}s_{23}s_{13}e^{i\delta_{13}}& s_{23}c_{13} \cr
s_{12}s_{23}-c_{12}c_{23}s_{13}e^{i\delta_{13}} & -c_{12}s_{23}-s_{12}c_{23}s_{13}e^{i\delta_{13}} & c_{23}c_{13} \cr
}
\end{equation}
where $c_{ij}=\cos\theta_{ij}$ and $s_{ij}=\sin\theta_{ij}$.\\

The range of values of the experimentally determined moduli in $|V^{exp}_{ij}|$, as given by Caso {\it et ~al} \cite{ref:3}, corresponds to $90\%$ confidence limits on the range of values of the mixing angles of

\begin{equation}\label{eq:4433}
0.217\leq s_{12}\leq 0.222,
\end{equation}

\begin{equation}\label{eq:4435}
0.036 \leq s_{23} \leq 0.042,
\end{equation}

\begin{equation}\label{eq:4437}
0.0018 \leq s_{13} \leq 0.0044.
\end{equation}

Seven of the nine absolute values of the CKM entries have been measured directly, by tree level processes. A range of values for the four parameters, $s_{12}, s_{23}, s_{13}$ and $\delta_{13}$ which is consistent with the seven direct measurements and the experimentally determined values of the moduli of $|{\bf V}|^{exp}$ \cite{ref:3}, is given by Nir \cite{ref:35}

\begin{equation}\label{eq:4438}
0.2173\leq s_{12}\leq 0.2219,
\end{equation}

\begin{equation}\label{eq:4441}
0.0378 \leq s_{23} \leq 0.0412,
\end{equation}

\begin{equation}\label{eq:4443}
0.00237 \leq s_{13} \leq 0.00395,
\end{equation}
$c_{13}$ is known to deviate from unity only in the sixth decimal place [\cite{ref:3}, \cite{ref:35}].\\

 The CP violating phase $\delta_{13}$, at present, is not constrained by direct measurements. However, the measurements of CP violation in $K$ decays \cite{ref:36} force $\delta_{13}$ to lie in the range

\begin{equation}\label{eq:4439}
0 \leq \delta_{13}\leq \pi.
\end{equation}

The standard parametrization ${\bf V}^{PDG}$ was introduced without taking the possible functional relations between the quark masses and the flavour mixing parameters into account. In contrast, these functional relations are explicitly exhibited in the theoretical expressions,  ${V}^{th}_{ij}$, derived in the previous sections. Furthermore, we have seen that, when the best values of the parameters $Z^{1/2}$ and $\Phi$ are used, the mixing matrix ${\bf V}^{th}$ reproduces the central values of all experimentally determined quantities, that is, the moduli $|V^{exp.}_{ij}|$, the Jarlskog invariant $J^{exp.}$ and the three inner  angles, $\alpha$, $\beta$ and $\gamma$, of the unitarity triangle \cite{ref:1}. Since the two parametrizations reproduce the same set of experimental data equally well, we are justified in writing 
\begin{equation}\label{eq:4311}
|V^{th}_{ij}|=|V^{PDG}_{ij}|=|V^{exp}_{ij}|.
\end{equation}
We cannot simply equate ${\bf V}^{th}$ and ${\bf V}^{PDG}$ because the arguments of corresponding matrix elements in the two parametrizations are not equal
\begin{equation}\label{eq:4312}
arg(V^{th}_{ij})\ne arg(V^{PDG}_{ij}).
\end{equation}
This difference is of no physical consequence, it reflects the freedom in choosing the unobservable phases of the quark fields in the mass representation. In the following, we will take advantage of this freedom to derive a phase transformed, theoretical mixing matrix, $\tilde {\bf V}^{th}$, related to ${\bf V}^{th}$ by a biunitary phase transformation, such that all corresponding entries in $\tilde {\bf V}^{th}$ and ${\bf V}^{PDG}$ are equal in modulus and phase. 
We will also derive exact, explicit expressions for the phases of the matrix elements $V^{PDG}_{ij}$ in terms of the phases of the matrix elements  $V^{th}_{ij}$, which, togheter with Eq. (\ref{eq:4311}), will be enough to show that ${\bf V}^{PDG}$ may be obtained from   ${\bf V}^{th}$ by means of a suitable rephasing of the quark fields in the mass representation.  
\subsection{Phase relations }
\label{sec:4.1}
In the mass basis, the quark charged currents take the form
\begin{equation}\label{eq:41}
{\it J}^{\mu}_c=\frac{\it g}{\sqrt 2}\bar q^{u}_{Li}\gamma^{\mu}V^{th}_{ij}q^{d}_{Lj}.
\end{equation}
A redefinition of the phases of the quark fields which leaves ${\it J}^{\mu}_c$
invariant, will change the argument of $V^{th}_{ij}$ but leave the moduli $|V^{th}_{ij}|$ invariant,
\begin{eqnarray}\label{eq:43}
V^{th}_{ij}\rightarrow \tilde V^{th}_{ij}=e^{-i\chi^u_i}V^{th}_{ij}e^{i\chi^d_j}.
\end{eqnarray}

The phases $\chi^u_i$ and $\chi^d_j$ ocurring in Eq. (\ref{eq:43}) will be determined from the requirement that corresponding  entries in $\tilde{\bf V}^{th.}$ and ${\bf V}^{PDG}$ be equal,

\begin{equation}\label{eq:45}
 |V^{th}_{ij}|e^{i(w^{th}_{ij}-(\chi^u_i-\chi^d_j))}=|V^{PDG}_{ij}|e^{iw^{PDG}_{ij}},
\end{equation}
in this expression $w^{th}_{ij}$ and $w^{PDG}_{ij}$ are the arguments of $V^{th}_{ij}$ and $V^{PDG}_{ij}$ respectively. Since the moduli $|V^{th}_{ij}|$ and $|V^{PDG}_{ij}|$ are equal, the arguments of the entries in the two parametrizations are related by the set of nine equations 
\begin{equation}\label{eq:47}
\chi^u_i-\chi^d_j=w^{th}_{ij}-w^{PDG}_{ij}.
\end{equation}

The set of Eqs.~(\ref{eq:47}) relate the differences of the unobservable quark field phases to the differences of the arguments of corresponding entries in ${\bf V}^{th}$ and ${\bf V}^{PDG}$. These two parametrizations of the mixing matrix are representations of the same set of experimental data. Therefore, it should be possible to derive, from Eqs.~(\ref{eq:47}), a new set of equations, expressing the five non-vanishing arguments $w^{PDG}_{ij}$ of $V^{PDG}_{ij}$ in terms only of the arguments $w^{th}_{ij}$ of $V^{th}_{ij}$ without making reference to the unobservable phases of the quark fields. With this purpose in mind, we notice that, in the left hand side of Eqs.~(\ref{eq:47}), there are nine differences of unobservables phases $\left( \chi^{(u)}_i-\chi^{(d)}_j\right)$, formed from only six different quark field phases. Differences of phases of the same quark field type, say $\left( \chi^{(d)}_j-\chi^{(d)}_{j'}\right)$, may be computed from Eqs.~(\ref{eq:47}) in at least three different ways. This redundancy implies the existence of non-trivial relations among the arguments of the entries of the two parametrizations. For example, from Eqs.~(\ref{eq:47}), the difference $\left( \chi^{(u)}_2-\chi^{(d)}_3\right)- \left( \chi^{(u)}_2-\chi^{(d)}_2\right)$ gives
\begin{equation}\label{eq:425}
\chi^{(d)}_2-\chi^{(d)}_3=w^{th}_{23}-w^{th}_{22}+w^{PDG}_{22},
\end{equation}
and the difference $\left( \chi^{(u)}_1-\chi^{(d)}_3\right)- \left( \chi^{(u)}_1-\chi^{(d)}_2\right)$ gives
\begin{equation}\label{eq:427}
\chi^{(d)}_2-\chi^{(d)}_3=w^{th}_{13}-w^{th}_{12}+\delta_{13}.
\end{equation}
If the phase difference  $(\chi^{(d)}_2-\chi^{(d)}_3)$ is eliminated between Eqs. (\ref{eq:425}) and (\ref{eq:427}) we get
\begin{equation}\label{eq:429}
\delta_{13}-w^{PDG}_{22}=w^{th}_{12}-w^{th}_{13}-w^{th}_{22}+w^{th}_{23}.
\end{equation}
Using the same elimination procedure for all possible combinations $\left( \chi^{(u)}_i-\chi^{(d)}_{j}\right)-\left( \chi^{(u)}_i-\chi^{(d)}_{j'}\right)$ we derive a set of nine equations, only four of which are linearly independent. 
One of these is Eq. (\ref{eq:429}), for the other three we may take
\begin{equation}\label{eq:431}
-w^{PDG}_{21}+w^{PDG}_{22}=w^{th}_{11}-w^{th}_{12}-w^{th}_{21}+w^{th}_{22},
\end{equation}

\begin{equation}\label{eq:433}
w^{PDG}_{31}-w^{PDG}_{32}=-w^{th}_{11}+w^{th}_{12}+w^{th}_{31}-w^{th}_{32},
\end{equation}
and
\begin{equation}\label{eq:435}
-w^{PDG}_{22}+w^{PDG}_{32}=-w^{th}_{22}+w^{th}_{23}+w^{th}_{32}-w^{th}_{33}.
\end{equation}

Since, in ${\bf V}^{PDG}$ there are five entries with non-vanishing arguments, namely, ~$w^{PDG}_{13}=-\delta_{13},  w^{PDG}_{21}, w^{PDG}_{22}, w^{PDG}_{31}$ and  $w^{PDG}_{32}$, we require still one more equation relating the arguments of the entries of the two parametrizations. This is obtained from the phase relations between the determinants of the two matrices,  ${\bf V}^{th}$  and ${\bf V}^{PDG}$. From Eqs.~(\ref{eq:43}) and (\ref{eq:45}), it follows that

\begin{eqnarray}\label{eq:49}
\det{\bf V}^{th}=\det\left[ {\bf X}^\dagger_u {\bf V}^{PDG}{\bf X}_d \right],
\end{eqnarray} 
in this expression ${\bf X}_u$ and ${\bf X}_d$ are the diagonal unitary matrices of phases ocurring in Eq. (\ref{eq:43}). The determinant of ${\bf V}^{PDG}$ is one, hence,
\begin{equation}\label{eq:411}
\det\left[{\bf X}^\dagger_u {\bf V}^{PDG}{\bf X}_d \right]=e^{i\sum^3_{i=1}\left(\chi^{(u)}_i-\chi^{(d)}_i\right)}.
\end{equation}
Similarly, from the definition of ${\bf V}^{th}$, Eq. (\ref{eq:37}), we get
\begin{equation}\label{eq:413}
\det {\bf V}^{th}=\det\left[ {\bf O}^T_u{\bf P}^{u-d}{\bf O}_d\right]=\det\left({\bf O}^T_u{\bf O}_d\right)\det{\bf P}^{u-d},
\end{equation}
the determinant of the orthogonal matrices is one, and the determinant of the diagonal matrix of phases ${\bf P}^{u-d}$ is $e^{i2\Phi}$. Taking for $\Phi$ the best value $\Phi^*=\pi/2$, we obtain

\begin{equation}\label{eq:415}
\det{\bf V}^{th}=e^{i2\Phi^*}=e^{i\pi}.
\end{equation}
 Substitution of Eq. (\ref{eq:411}) and Eq. (\ref{eq:415}) in Eq. (\ref{eq:49}) gives
\begin{equation}\label{eq:417}
\sum^{3}_{i=1}\left( \chi^{(u)}_i-\chi^{(d)}_i\right)=2\Phi^*=\pi.
\end{equation}
This phase relation guarantees the equality of the determinants of $\tilde{\bf V}^{th}$ and ${\bf V}^{PDG}$.\\
The sum of the unobservable quark field phases ocurring in the left hand side of Eq.~(\ref{eq:417}) may be computed from Eqs.~(\ref{eq:47}),
\begin{equation}\label{eq:419}
\sum^{3}_{i=1}\left( \chi^{(u)}_i-\chi^{(d)}_i\right)=\sum^{3}_{i=1}w^{th}_{ii}-w^{PDG}_{22}.
\end{equation}
Now, we eliminate the unobservable quark field phases between Eq.~(\ref{eq:417}) and Eq. (\ref{eq:419}), to get,
\begin{equation}\label{eq:423}
w^{PDG}_{22}=\sum^{3}_{i=1}w^{th}_{ii}-2\Phi^*.
\end{equation}
This,  relation shows that $\arg(V^{PDG}_{22})$ is uniquely determined $(mod ~2\pi)$ in terms of the arguments of the entries in ${\bf V}^{th}$.\\
 
With the help of Eq. (\ref{eq:423}) we solve Eqs. (\ref{eq:429})-(\ref{eq:435}) for all the other non-vanishing arguments of ${\bf V}^{PDG}$
\begin{equation}\label{eq:437}
\delta_{13}=w^{th}_{11}+w^{th}_{12}-w^{th}_{13}+w^{th}_{23}+w^{th}_{33}-2\Phi^*
\end{equation}

\begin{equation}\label{eq:439}
w^{PDG}_{21}=w^{th}_{21}+w^{th}_{12}+w^{th}_{33}-2\Phi^*
\end{equation}

\begin{equation}\label{eq:441}
w^{PDG}_{31}=w^{th}_{31}+w^{th}_{12}+w^{th}_{23}-2\Phi^*
\end{equation}

\begin{equation}\label{eq:443}
w^{PDG}_{32}=w^{th}_{32}+w^{th}_{23}+w^{th}_{11}-2\Phi^*.
\end{equation}
In this way, we have shown that the arguments $w^{PDG}_{ij}$ of $V^{PDG}_{ij}$ are uniquely determined (mod ~$2\pi$) by the arguments $w^{th}_{ij}$ of $V^{th}_{ij}$.\\

We now return to the question of the quark field phases and the phase transformation from $V^{th}_{ij}$ to $V^{PDG}_{ij}$.
Substitution of Eqs.~(\ref{eq:423})-(\ref{eq:443}) into Eqs.~(\ref{eq:47}), gives the differences of the quark field phases explicitly in terms of the known arguments $w^{th}_{ij}$ of $V^{th}_{ij}$.   The quark field phases themselves are determined only up to a common additive constant. Since the quark field phases are unobservable, without loss of generality, we may fix one of them, and solve for the others. In this way, if we set $\chi^{d}_{1}=0$, we get
\begin{eqnarray}\label{eq:445}
\chi^{d}_{1}&=&0^\circ,\cr
\chi^{d}_{2}&=&w^{th}_{11}-w^{th}_{12},\cr
\chi^{d}_{3}&=&-w^{th}_{23}-w^{th}_{33}-w^{th}_{12}+2\Phi^*,\cr
\chi^{u}_{1}&=&w^{th}_{11},\cr
\chi^{u}_{2}&=&-w^{th}_{12}-w^{th}_{33}+2\Phi^*,\cr
\chi^{u}_{3}&=&-w^{th}_{23}-w^{th}_{12}+2\Phi^*.
\end{eqnarray}
Then, the diagonal matrices of phases required to compute the phase transformed $\tilde {\bf V}^{th}$ are
\begin{equation}\label{eq:447}
{\bf X}_u=diag[e^{iw^{th}_{11}}, e^{i(-w^{th}_{12}-w^{th}_{33}+2\Phi^*)}, e^{i(-w^{th}_{23}-w^{th}_{12}+2\Phi^*)}]
\end{equation}
and
\begin{equation}\label{eq:449}
{\bf X}_d=diag[1, e^{i(w^{th}_{11}-w^{th}_{12})}, e^{i(w^{th}_{12}-w^{th}_{23}-w^{th}_{33}+2\Phi^*)}].
\end{equation}
Hence, with the help of Eqs. (\ref{eq:437})-(\ref{eq:443}), we verify that
\begin{equation}\label{eq:451}
{\bf X^{\dagger}_u}{\bf V}^{th}{\bf X_d}={\bf V}^{PDG}
\end{equation}
is satisfied as an identity, provided that $|V^{th}_{ij}|=|V^{PDG}_{ij}|$.
\section{The mixing angles}
\label{sec:5}
The invariant measurables of the quark mixing matrix are the moduli of its elements i.e., the quantities $|V_{ij}|$, and the Jarlskog invariant $J$. But even $J$, up to a sign, is a function of the moduli \cite{ref:4}. Hence, two different parametrizations, are equivalent if the moduli of the corresponding entries are equal. In the case of $V^{th}_{ij}$ and $V^{PDG}_{ij}$, when the best set of adjustable parameters of each parametrization, $(Z^{1/2},~\Phi)$ and $(\theta_{12},~\theta_{23},~\theta_{13},~\delta_{13})$ respectively, is used to fit the experimental data, the moduli of corresponding entries of the two parametrizations are numerically equal and give an equally good representation of the experimentally determined values of the moduli of the mixing matrix $|V^{exp}_{ij}|$ \cite{ref:3}. Therefore we are justified in writing
\begin{equation}\label{eq:51}
|V^{th}_{ij}|=|V^{PDG}_{ij}|,
\end{equation}
even though $V^{th}_{ij}$ has only two adjustable parameters $(Z^{1/2},~\Phi)$ while the number of adjustable parameters in $V^{PDG}_{ij}$ is four, namely, $(\theta_{12},~\theta_{23},~\theta_{13},~\delta_{13})$.
All entries in $|V^{th}_{ij}|$ are explicit functions of the four quark mass ratios $(m_{u}/m_{t}, m_{c}/m_{t}, m_{d}/m_{b}, m_{s}/m_{b})$ and the two parameters $Z^{1/2}$ and $\Phi$. The equality of the moduli of corresponding entries of the two parametrizations will allow us to derive explicit expressions for the mixing angles in terms of the four quark mass ratios $(m_{u}/m_{t}, m_{c}/m_{t}, m_{d}/m_{b}, m_{s}/m_{b})$ and the parameters $Z^{1/2}$ and $\Phi$.

From the equality of $|V^{th}_{13}|$ and $|V^{PDG}_{13}|$, it follows that

\begin{equation}\label{eq:53}
\sin\theta_{13}=|V^{th}_{ub}|,
\end{equation}
if we take $|V^{th}_{ub}|$ from (\ref{eq:327}), and we set $\Phi$ and $Z^{1/2}$ equal to their best values $\Phi^*=\pi/2$ and $Z^{1/2*}=\sqrt{\frac{81}{32}}$, we get
\begin{eqnarray}\label{eq:55}
\sin\theta_{13}&=&\bigg\{ \frac{\tilde m_{c}(1-\tilde m_{u}-\delta^{*}_{u})\tilde m_{d}\tilde m_{s}\delta^{*}_{d}}{(1-\delta^*_{u})(1-\tilde m_{u})(\tilde m_{c}+\tilde m_{u})(1-\delta^*_{d})(1+\tilde m_{s})(1-\tilde m_{d})}\cr
&+& \bigg [ \left( \frac{\tilde m_{u}(1-\tilde m_{u}-\delta^{*}_{u})\delta^{*}_{d}}{(1-\tilde m_{u})(\tilde m_{c}+\tilde m_{u})(1+\tilde m_{s})(1-\tilde m_{d})}\right)^{1/2}\cr
&-&\left(\frac{\tilde m_{u}(1+\tilde m_{c}-\delta^{*}_u)\delta^{*}_u(1-\tilde m_{d}-\delta^*_d)(1+\tilde m_s-\delta^*_d)}{(1-\delta^*_u)(1-\tilde m_u)(\tilde m_c +\tilde m_u)(1-\delta^*_d)(1+\tilde m_s)(1-\tilde m_d)}\right)^{1/2}
\bigg ]^{2} \bigg\}^{\frac{1}{2}}
\end{eqnarray}
The computation of $\sin\theta_{23}$ is slightly more involved. From Eq. (\ref{eq:4431}) and the equality of  $|V^{th}_{ij}|$ and $|V^{PDG}_{ij}|$, we obtain

\begin{eqnarray}\label{eq:57}
\sin\theta_{23}=\frac{|V^{PDG}_{cb}|}{\sqrt{1-|V^{PDG}_{ub}|^{2}}}=\frac{|V^{th}_{cb}|}{\sqrt{1-|V^{th}_{ub}|^{2}}}.
\end{eqnarray}
Substitution of the expressions (\ref{eq:331}) and (\ref{eq:327}) with $\Phi^*=\pi/2$ and $Z^{*1/2}=\sqrt\frac{81}{32}$ for $|V^{th}_{cb}|$ and $|V^{th}_{ub}|$ in Eq. (\ref{eq:57}) gives
\begin{eqnarray}\label{eq:59}
\sin\theta_{23}&=&\sqrt{\frac{1-\tilde m_u}{1+\tilde m_c}} 
\bigg\{
\tilde m_u(1+\tilde m_c-\delta^{*}_u)\tilde m_d\tilde m_s\delta^{*}_d+
\bigg [\sqrt{ (1-\delta^*_u)\tilde m_c(1+\tilde m_c-\delta^*_u)(1-\delta^*_d)\delta^*_d}\cr &-&
\sqrt{\tilde m_c(1-\tilde m_u-\delta^*_u)\delta^*_u(1-\tilde m_d-\delta^*_d)(1+\tilde m_s-\delta^*_d)}\bigg ]^2\bigg\}^{1/2}
\cr &\times&
\bigg\{ -\bigg[\sqrt{(1-\delta^*_u)\tilde m_u (1-\tilde m_u-\delta^*_u)(1-\delta^*_d)\delta^*_d}\cr &-&
\sqrt{\tilde m_u(1+\tilde m_c-\delta^*_u)\delta^*_u(1-\tilde m_d-\delta^*_d)(1+\tilde m_s-\delta^*_d)}\bigg]^2 \cr&+&
(1-\delta^*_u)(1-\tilde m_u)(\tilde m_c+\tilde m_u)(1-\delta^*_d)(1+\tilde m_s)(1-\tilde m_d)-\tilde m_c(1-\tilde m_u-\delta^*_u)\tilde m_d\tilde m_s\delta^*_d
\bigg\}^{-1/2}.
\end{eqnarray}
Similarly from Eq. (\ref{eq:4431}) and the equality of $|V^{th}_{12}|$ and $|V^{PDG}_{12}|$, we obtain
\begin{eqnarray}\label{eq:511}
\sin\theta_{12}=\frac{|V^{PDG}_{us}|}{\sqrt{1-|V^{PDG}_{ub}|^{2}}}=\frac{|V^{th}_{us}|}{\sqrt{1-|V^{th}_{ub}|^{2}}}.
\end{eqnarray}
Then, substitution of the expressions (\ref{eq:325}) and (\ref{eq:327}) for $|V^{th}_{us}|$ and $|V^{th}_{ub}|$ in (\ref{eq:511}) gives
\begin{eqnarray}\label{eq:513}
\sin\theta_{12}&=&\sqrt{\frac{1-\tilde m_d}{\tilde m_s+\tilde m_d}} 
\bigg\{
\tilde m_c(1-\tilde m_u-\delta^{*}_u) \tilde m_d(1+\tilde m_s-\delta^{*}_d)
\cr &+&
\bigg [\sqrt{ (1-\delta^*_u)\tilde m_u(1-\tilde m_u-\delta^*_u)(1-\delta^*_d)\tilde m_s(1+\tilde m_s-\delta^*_d)}\cr &+&\sqrt{\tilde m_u(1+\tilde m_c-\delta^*_u)\delta^*_u\tilde m_s(1-\tilde m_d-\delta^*_d)\delta^*_d}\bigg ]^2\bigg\}^{1/2}\cr &\times&
\bigg\{-\bigg[\sqrt{(1-\delta^*_u)\tilde m_u (1-\tilde m_u-\delta^*_u)(1-\delta^*_d)\delta^*_d}\cr&-&
\sqrt{\tilde m_u(1+\tilde m_c-\delta^*_u)\delta^*_u(1-\tilde m_d-\delta^*_d)(1+\tilde m_s-\delta^*_d)}\bigg]^2\cr&+&
(1-\delta^*_u)(1-\tilde m_u)(\tilde m_c+\tilde m_u)(1-\delta^*_d)(1+\tilde m_s)(1-\tilde m_d)\cr &-&
\tilde m_c(1-\tilde m_u-\delta^*_u)\tilde m_d\tilde m_s\delta^*_d \bigg\}^{-1/2}.
\end{eqnarray}
The computed values for $\sin\theta_{12},\sin\theta_{23}$ and $\sin\theta_{13}$ corresponding to the best $\chi^2$ fit of $|V^{th}_{ij}|$, $J^{th}$ and $\alpha^{th}$, $\beta^{th}$ and $\gamma^{th}$ to the experimentally determined quantities $|V^{exp}_{ij}|$, $J^{exp}$ and the three inner angles of the unitarity triangle $\alpha^{exp}$, $\beta^{exp}$ and $\gamma^{exp}$ are obtained when the numerical values of $|V^{th}_{us}|$, $|V^{th}_{ub}|$ and $|V^{th}_{cb}|$ computed from Eqs.(\ref{eq:325}), (\ref{eq:327}), (\ref{eq:331}) and given in Eq.~(\ref{eq:3.21}) are substituted in to Eqs.~(\ref{eq:53}), (\ref{eq:57}) and (\ref{eq:511}). In this way, we get
\begin{eqnarray}\label{eq:515}
\sin\theta^*_{12}=0.221,
\end{eqnarray}
\begin{eqnarray}\label{eq:517}
\sin\theta^*_{23}=0.040,
\end{eqnarray}
\begin{eqnarray}\label{eq:519}
\sin\theta^*_{13}=0.0034.
\end{eqnarray}
The numerical value of $cos\theta^*_{13}$ deviates from unity in the sixth decimal place.\\

We notice that the numerical values of the mixing angles computed from quark masses and the best values of the symmetry breaking parameters coincide almost exactly with the central values of the experimentally determined quantities, as could be expected from Eq.~(\ref{eq:51}). This observation is interesting because, in the case of three families, the most general form of the mixing matrix has at most four free, independent parameters \cite{ref:4} which could be four independent moduli or three mixing angles and one phase as occurs in  ${\bf V}^{PDG}$. The symmetry derived ${\bf V}^{th}$ has only two free, real independent parameters. In spite of that, the quality of the fit of ${\bf V}^{th}$ to the experimental data is as good as the quality of the fit of ${\bf V}^{PDG}$ to the same data. The predictive power of ${\bf V}^{th}$ implied by this fact originates in the flavour permutational symmetry of the Standard Model and the assumed symmetry breaking pattern  from which the texture in the quark mass matrices and ${\bf V}^{th}$ were derived. 
\section{The Cp violating phase $\delta_{13}$}
\label{sec:6}
The CP violating phase $\delta_{13}$ of the standard parametrization ${\bf V}^{PDG}$ of the quark mixing matrix is given in Eq. (\ref{eq:437}) in terms of the arguments $w^{th}_{ij}$ of five entries in the theoretical expression for  $V^{th}_{ij}$ and the corresponding CP violating phase $\Phi$. Taking from Eq. (\ref{eq:3.21}) the numerical values of the arguments  $w^{th}_{ij}$ and setting $\Phi$ equal to the best value $\Phi^*=\pi/2$, we obtain the numerical value of $\delta_{13}$ corresponding to the best fit of $|V^{th}_{ij}|$ to the experimental data
\begin{equation}\label{eq:61}
\delta^*_{13}=75^{\circ}.
\end{equation}
This predicted value of  $\delta_{13}$ is very close to the numerical value of the third inner angle $\gamma$, of the unitarity triangle. The difference may readily be computed in terms of the arguments  $w^{th}_{ij}$. From the expression for $\gamma$ 
\begin{equation}\label{eq:62}
\gamma=arg\left[ -\frac{V^{*}_{cb}V_{cd}}{V^{*}_{ub}V_{ud}}\right]
\end{equation}
we get
\begin{equation}\label{eq:63}
-\gamma=w^{th}_{11}-w^{th}_{13}-w^{th}_{21}+w^{th}_{23}+\pi
\end{equation}
which, when compared with the expression (\ref{eq:437}) for $\delta_{13}$ gives
\begin{equation}\label{eq:65}
-\gamma=\delta_{13}-(w^{th}_{12}+w^{th}_{21}+w^{th}_{33}-2\Phi^*-\pi).
\end{equation}
Taking from Eq. (\ref{eq:3.21}) the numerical values of the arguments corresponding to the best values $\Phi^*=90^{\circ}$ and $Z^{1/2}*= \sqrt{\frac{81}{32}}$, we obtain
\begin{equation}\label{eq:67}
(w^{th}_{12}+w^{th}_{21}+w^{th}_{33}-2\Phi^*-\pi)=0.04^{\circ}.
\end{equation}
This is, indeed, a very small number, and justifies the approximation
\begin{equation}\label{eq:69}
-\gamma \approx \delta^*_{13}.
\end{equation}
According to this, the value of $|\gamma|$ computed from quark mass ratios and the best values of the parameters $Z^{*1/2}$ and $\Phi^*$ is $|\gamma|=75^{\circ}$, in agreement with the bounds extracted from the precise measurements of the $B^0_d$ oscillation frequency \cite{ref:32} and the measurements of the rates of the exclusive hadronic decays $B^{\pm}\rightarrow \pi K$ and the CP averaged $B^{\pm}\rightarrow \pi^{\pm}\pi^0$ \cite{ref:37}.
Exact explicit expressions for the CP violating phase $\delta_{13}$ in terms of the four quark mass ratios and the parameters $Z^{*1/2}$ and $\Phi^*$ may readily be found; such an expression could be derived from Eq. (\ref{eq:437}) in terms of the arguments of five matrix elements of ${\bf V}^{th}$. However, a simpler expression, involving only four matrix elements of ${\bf V}^{th}$ may be obtained from the Jarlskog invariant $J$. \\

   The Jarlskog invariant may be written in terms of four matrix elements of ${\bf V}$ as
\begin{equation}\label{eq:611}
J={\it Im}[V_{12}V_{23}V^*_{13}V^*_{22}].
\end{equation}
Since $J$ is an invariant, its value is independent of the particular parametrization of ${\bf V}$. If we write the right hand side of Eq. (\ref{eq:611}) in terms of the standard parametrization ${\bf V}^{PDG}$, we obtain

\begin{equation}\label{eq:613}
\sin\delta_{13}=\frac{J^{th}}{s_{12}s_{13}s_{23}c_{12}c^2_{13}c_{23}}.
\end{equation} 
The terms in the denominator in the right hand side of this expression were written in Eqs. (\ref{eq:53}), (\ref{eq:57}) and (\ref{eq:511}) in terms of the moduli $|V^{th}_{12}|$, $|V^{th}_{13}|$ and $|V^{th}_{23}|$. Hence,
\begin{equation}\label{eq:615}
s_{12}s_{13}s_{23}c_{12}c^2_{13}c_{23}=\frac{|V^{th}_{12}||V^{th}_{13}||V^{th}_{23}|[(1-|V^{th}_{13}|^2-|V^{th}_{12}|^2)(1-|V^{th}_{13}|^2-|V^{th}_{23}|^2)]^{1/2}}{1-|V^{th}_{13}|^2}.
\end{equation}
Substitution of Eq. (\ref{eq:615})  in Eq. (\ref{eq:613}) gives
\begin{equation}\label{eq:617}
\sin{\delta_{13}}=\frac{(1-|V^{th}_{13}|^2){\it Im}[V^{th}_{12}V^{th}_{23}V^{th*}_{13}V^{th*}_{22}]}
{|V^{th}_{12}||V^{th}_{13}||V^{th}_{23}|\sqrt{ (1-|V^{th}_{13}|^2-|V^{th}_{12}|^2)(1-|V^{th}_{13}|^2-|V^{th}_{23}|^2)}},
\end{equation}
the right hand side of this equation may be written in terms of the quark mass ratios and the symmetry breaking parameters $Z^{*1/2}$ and $\Phi^*$ with the help of Eqs.  (\ref{eq:55}), (\ref{eq:59}) and (\ref{eq:513}).\\
A simpler expression which leads to a very accurate approximation for $\delta_{13}$ is obtained from Eq. (\ref{eq:617}) if the matrix elements in the square brackets are written as modulus and argument, and use is made of the unitarity of ${\bf V}^{th}$ to simplify the denominator,
\begin{equation}\label{eq:619}
\sin{\delta_{13}}=\frac{(1-|V^{th}_{13}|^2)|V^{th}_{22}| \sin{(w^{th}_{12}+w^{th}_{23}-w^{th}_{13}-w^{th}_{22})}}
{ |V^{th}_{11}||V^{th}_{33}|}.
\end{equation}
Explicit expressions for the arguments $w^{th}_{12}, w^{th}_{23}, w^{th}_{13}$ and $w^{th}_{22}$ in terms of the quark mass ratios may be derived from  Eqs. (\ref{eq:325})-(\ref{eq:331}) setting $Z^{1/2}$ and $\Phi$ equal to their best values $Z^{*1/2}=\sqrt{\frac{81}{32}}$ and $\Phi^*=\pi/2$, we get
\begin{equation}\label{eq:621}
w^{th}_{us}=\pi-\tan^{-1}\left( \sqrt{ \frac{\tilde m_u\tilde m_s}{\tilde m_c\tilde m_d}}\left[\sqrt{(1-\delta^*_u)(1-\delta^*_d)}+\sqrt{ \delta^*_u\delta^*_d\frac{(1+\tilde m_c-\delta^*_u)(1-\tilde m_d-\delta^*_d)}{(1-\tilde m_u-\delta^*_u)(1+\tilde m_s-\delta^*_d)}}\right]\right),
\end{equation}

\begin{equation}\label{eq:623}
w^{th}_{cb}=\pi-\tan^{-1}\left( \sqrt{ \frac{\tilde m_c}{\tilde m_u\tilde m_d\tilde m_s}}\left[\sqrt{(1-\delta^*_u)(1-\delta^*_d)}-\sqrt{\frac{\delta^*_u(1-\tilde m_u-\delta^*_u)(1-\tilde m_d-\delta^*_d)(1+\tilde m_s-\delta^*_d)}{\delta^*_d(1+\tilde m_c-\delta^*_u)}}\right]\right),
\end{equation}

\begin{equation}\label{eq:625}
w^{th}_{ub}=\tan^{-1}\left( \sqrt{ \frac{\tilde m_u}{\tilde m_c\tilde m_d\tilde m_s}}\left[\sqrt{(1-\delta^*_u)(1-\delta^*_d)}-\sqrt{\frac{\delta^*_u(1+\tilde m_c-\delta^*_u)(1-\tilde m_d-\delta^*_d)(1+\tilde m_s-\delta^*_d)}{\delta^*_d(1-\tilde m_u-\delta^*_u)}}\right]\right),
\end{equation}

\begin{equation}\label{eq:627}
w^{th}_{cs}=\tan^{-1}\left( \sqrt{ \frac{\tilde m_c\tilde m_s}{\tilde m_u\tilde m_d}}\left[\sqrt{(1-\delta^*_u)(1-\delta^*_d)}+\sqrt{ \delta^*_u\delta^*_d\frac{(1-\tilde m_u-\delta^*_u)(1-\tilde m_d-\delta^*_d)}{(1+\tilde m_c-\delta^*_u)(1+\tilde m_s-\delta^*_d)}}\right]\right).
\end{equation}

Computing the second factor in square brackets in the leading order of magnitude, we get

\begin{equation}\label{eq:629}
w^{th}_{us}\approx\pi-\tan^{-1}\left( \sqrt{ \frac{\tilde m_u\tilde m_s}{\tilde m_c\tilde m_d}}\right),
\end{equation}

\begin{equation}\label{eq:631}
w^{th}_{cb}\approx\pi-\tan^{-1}\left( \sqrt{ \frac{\tilde m_c}{\tilde m_u\tilde m_d\tilde m_s}}\left[(1-\sqrt{\frac{\delta^*_u}{\delta^*_d}})\right]\right),
\end{equation}

\begin{equation}\label{eq:633}
w^{th.}_{ub}\approx\tan^{-1}\left( \sqrt{ \frac{\tilde m_u}{\tilde m_c\tilde m_d\tilde m_s}}\left[(1-\sqrt{\frac{\delta^*_u}{\delta^*_d}})\right]\right),
\end{equation}
and
\begin{equation}\label{eq:635}
w^{th}_{cs}\approx\tan^{-1}\left( \sqrt{ \frac{\tilde m_c\tilde m_s}{\tilde m_u\tilde m_d}}\right).
\end{equation}
 The modulus $|V^{th}_{ub}|$ has already been expressed in terms of quark mass ratios and the parameters characterizing the symmetry breaking pattern $Z^{*1/2}$ and $\Phi^*$, in Eqs. (\ref{eq:53}) and (\ref{eq:55}). Similar expressions for the other moduli ocurring in Eq. (\ref{eq:619}) may also be given

\begin{eqnarray}\label{eq:637}
|V_{ud}|&=&\left(\frac{\tilde {m}_c \left( 1-\tilde{m}_u -\delta^*_u \right)
\tilde {m}_s \left( 1-\tilde{m}_d -\delta^*_d \right)}{\left( 1-\delta^*_u \right)\left( 1-\tilde{m}_u \right)\left( \tilde {m}_c+\tilde {m}_u \right)
\left( 1-\delta^*_d \right)\left( 1-\tilde{m}_d \right)
\left( \tilde {m}_s+\tilde {m}_d \right)}\right)^{1/2}\cr &\times&\left\{1+
\frac{\tilde{m}_u \tilde{m}_d}{\tilde{m}_c \tilde{m}_s}\left[\left((1-\delta^*_u)(1-\delta^*_d) \right)^{1/2}+\left( \delta^*_u\delta^*_d\frac{(1+\tilde {m}_c-\delta^*_u)(1+\tilde {m}_s-\delta^*_d)}{(1-\tilde {m}_u-\delta^*_u)(1-\tilde {m}_d-\delta^*_d)}\right)^{1/2}\right]^2\right\},
\end{eqnarray}

\begin{eqnarray}\label{eq:639}
|V_{cs}|&=&\left(\frac{\tilde {m}_c
\left( 1+\tilde{m}_c -\delta^*_u \right)\tilde {m}_s
\left( 1+\tilde{m}_s -\delta^*_d \right)}{
\left( 1+\tilde{m}_c \right)\left( \tilde {m}_c+\tilde {m}_u \right)
\left( 1+\tilde{m}_s \right)
\left( \tilde {m}_s+\tilde {m}_d \right)}\right)^{1/2}\cr &\times&
\bigg\{\left[1+
\left(\frac{\delta^*_u\delta^*_d\left( 1-\tilde {m}_u-\delta^*_u \right)
\left( 1-\tilde{m}_d -\delta^*_d \right)}{\left( 1-\delta^*_u \right)\left( 1-\delta^*_d \right)
\left( 1+\tilde{m}_c-\tilde{m}_u \right)
\left( 1+\tilde{m}_s-\delta^*_d \right)}\right)^{1/2}\right]^2 \cr&+&
\frac{\tilde{m}_u \tilde{m}_d}{\tilde{m}_c \tilde{m}_s}\frac{1}{( 1-\delta^*_u )( 1-\delta^*_d)}
\bigg\}^{1/2},
\end{eqnarray}

\begin{eqnarray}\label{eq:641}
|V_{tb}|&=&\left[\frac{(1-\tilde{m}_u -\delta^*_u)(1+\tilde{m}_c -\delta^*_u)
(1-\tilde{m}_d -\delta^*_d)(1-\tilde m_s -\delta^*_d) }
{(1- \delta^*_u)(1+\tilde{m}_c)(1-\tilde{m}_u)
(1- \delta^*_d )(1+\tilde{m}_s)(1-\tilde{m}_d)}
\right]^{1/2} 
\cr &\times&\bigg\{\left[1+\left(\frac{\delta^*_u\delta^*_d(1-\delta^*_u)(1- \delta^*_d )}
{(1+\tilde{m}_c-\delta^*_u)(1-\tilde{m}_u-\delta^*_u)
(1-\tilde{m}_d-\delta^*_d)(1+\tilde{m}_s-\delta^*_d)}\right)^{1/2}\right]^2\cr
&+& \frac{\tilde {m}_u \tilde {m}_c \delta^*_u 
\tilde {m}_d \tilde{m}_s \delta^*_d  }
{(1-\tilde {m}_u- \delta^*_u )(1 + \tilde {m}_c- \delta^*_u )(1+ \tilde {m}_s- \delta^*_d)(1-\tilde{m}_d-\delta^*)}\bigg\}^{1/2}.
\end{eqnarray}

Computing in the leading order of magnitude, the first factor in the right hand side of Eq. (\ref{eq:619}) gives
\begin{equation}\label{eq:643}
\frac{(1-|V^{th}_{13}|^2)|V^{th}_{22}| }{ |V^{th}_{11}||V^{th}_{33}|}\approx \frac{(1-\delta^*_u)(1-\tilde m_u)(1-\delta^*_d)(1-\tilde m_d)}{(1-\tilde m_u-\delta^*_u)(1-\tilde m_d-\delta^*_d)}\left( 1-\frac{\tilde m_u}{\tilde m_d}(\sqrt{\delta^*_d}-\sqrt{\delta^*_u})\right).
\end{equation}
Inserting in to Eq.~(\ref{eq:643}) the numerical values of the mass ratios and $\sqrt{\delta^*_d}-\sqrt{\delta^*_u}=0.04$, we find that the right hand side of Eq.(\ref{eq:643}) differs from one in the third decimal place, 
\begin{equation}\label{eq:645}
\frac{(1-|V^{th}_{13}|^2)|V^{th}_{22}| }{ |V^{th}_{11}||V^{th}_{33}|}\approx 1.
\end{equation}
Therefore, 
\begin{equation}\label{eq:647}
\sin\delta^*_{13}\approx \sin(w^{th}_{uc}+w^{th}_{cb}-w^{th}_{ub}-w^{th}_{cs}),
\end{equation}
taking the numerical values of the argument in the right hand side of Eq. (\ref{eq:647}) from (\ref{eq:3.17}), we obtain
\begin{equation}\label{eq:649}
\delta^*_{13}\approx 75^{\circ},
\end{equation}
in agreement with Eq. (\ref{eq:61}).
The approximate expression Eq. (\ref{eq:647}) for $\sin\delta^*_{13}$ could also be derived from Eq. (\ref{eq:429}) if $w^{PDG}_{22}$ is neglected. Computing $w^{PDG}_{22}$ from Eq. (\ref{eq:423}) and (\ref{eq:621})-(\ref{eq:627}), we obtain $w^{PDG}_{22}=-0.0018^{\circ}$ which shows that Eq. (\ref{eq:647}) is a very good approximation. Since Eq. (\ref{eq:423}) was derived from the the phase relations expressing the arguments of $V^{PDG}_{ij}$  in terms of those of  $V^{th}_{ij}$, while Eq. (\ref{eq:619})  was derived from the expression Eq. (\ref{eq:617}) for the Jarlskog invariant, the agreement found between  Eqs. (\ref{eq:423}) and  (\ref{eq:619})- (\ref{eq:647}) is a consistency check of our formalism.
\section{Summary and Conclusions}
\label{sec:7}

In this work, we explicitly exhibit the phase equivalence of the theoretical mixing matrix, ${\bf V}^{th}$, derived from the breaking  of the flavour permutational symmetry in a previous work ~\cite{ref:1}, and the standard parametrization   ~\cite{ref:2}, ${\bf V}^{PDG}$, advocated by the Particle Data Group ~\cite{ref:3}.  More precisely, we show that when the best set of adjustable parameters of each parametrization is obtained from a $\chi^2$ fit to the same experimental data, the moduli of corresponding entries in the matrices are numerically equal and give an equally good representation of experimentally determined values of the moduli of the mixing matrix  ${\bf V}^{exp}$. Hence we are justified in writing
\begin{equation}\label{eq:70}
|{V}^{th}_{ij}|=|{V}^{PDG}_{ij}|
\end{equation}
even though ${\bf V}^{th}$ has only two adjustable parameters $Z^{1/2}$ and $\Phi$, while the number of adjustable parameters in ${\bf V}^{PDG}$ is four, the three mixing angles $\theta_{12}, \theta_{23}, \theta_{13}$ and the CP violating phase $\delta_{13}$. From this result, we proceed to fomulate and solve the equations of the rephasing transformation which acting on ${\bf V}^{th}$ gives a phase transformated  $\tilde{\bf V}^{th}$ such that the corresponding entries in $\tilde{\bf V}^{th}$ and ${\bf V}^{PDG}$ are equal in modulus and phase. As part of the solution, we obtain a set of equations expressing the non-vanishing arguments $w^{PDG}_{ij}$ of the matrix elements $V^{PDG}_{ij}$ of the standard parametrization in terms of the arguments $w^{th}_{ij}$ of the entries in the flavour symmetry derived $V^{th}_{ij}$. Since the matrix elements of ${\bf V}^{th}$ are known functions of the quark mass ratios and the parameters $Z^{1/2}$ and $\Phi$, we obtain exact, explicit, analytical expressions for the arguments $w^{PDG}_{ij}$ as functions of the quark mass ratios and the parameters $Z^{*1/2}$ and $\Phi^*$ which characterize the best or preferred symmetry breaking pattern. In particular, we derive an exact, explicit expression for the CP violating phase $\delta^*_{13}$ written as a linear combination of the arguments of five entries in  ${\bf V}^{th}$. Similarly, from the equality of the moduli of the corresponding entries in the two parametrizations, we solve for the mixing parameters $\sin\theta_{12}, ~\sin\theta_{13}, ~\sin\theta_{23}$, ocurring in the standard parametrization, in terms of the moduli  $|V^{th}_{ij}|$. Then, using the explicit expressions found for  $V^{th}_{ij}$ in our previous work \cite{ref:1}, we obtain exact, explicit expressions for the mixing parameters $\sin\theta^*_{12}, ~\sin\theta^*_{13}, ~\sin\theta^*_{23}$, in terms of the quark mass ratios and the parameters $Z^{*1/2}$ and $\Phi^*$. From these results and an expression for the Jarlskog invariant, written in terms of four matrix elements of ${\bf V}^{th}$, we derive an alternative, explicit, analytical expression for $\sin\delta^*_{13}$ as function of the quark mass ratios and the parameters $Z^{*1/2}$ and $\Phi^*$.\\

In conclusion, in the standard electroweak model of particle interactions, both the masses of the quarks as well as the mixing parameters and the CP violating phase appear as free, independent parameters. In this work we have given explicit expressions for the mixing parameters $\sin\theta_{12}, ~\sin\theta_{13}, ~\sin\theta_{23}$ and the CP violating phase $\delta_{13}$ of the standard parametrization of the mixing matrix \cite{ref:2} as functions of the four quark mass ratios $m_u/m_t$, $m_c/m_t$, $m_d/m_b$, $m_s/m_b$, and two parameters: $Z^{1/2}$ and $\Phi$. These expressions were obtained from a simple and explicit ansatz for the pattern of the breaking of quark flavour symmetry and a rephasing transformation of the quark fields in the mass representation.\\

The numerical values of the mixing parameters  $\sin\theta^*_{12}, ~\sin\theta^*_{13}$ and $\sin\theta^*_{23}$ computed from quark mass ratios and the best values of the parameters $Z^{*1/2}$ and $\Phi^*$, coincide almost exactly with the central values of the same mixing parameters, determined from the experimental data \cite{ref:35}, as could be expected from the phase equivalence of  ${\bf V}^{th}$ and  ${\bf V}^{PDG}$, expresed in Eq.~(\ref{eq:70}). This observation is interesting because, in the case of three families, the most general form of the mixing matrix has at most four free, independent parameters \cite{ref:4} which could be four independent moduli or three mixing angles and one phase as occurs in ${\bf V}^{PDG}$. The symmetry derived ${\bf V}^{th}$ has only two free, real independent parameters. In spite of that, the quality of the fit of  ${\bf V}^{th}$ to the experimental data is as good as the quality of the fit of ${\bf V}^{PDG}$ to the same data. The predictive power of ${\bf V}^{th}$ implied by this fact originates in the flavor permutational symmetry of the Standard Model and the assumed symmetry breaking pattern from which the texture in the quark mass matrices and ${\bf V}^{th}$ were derived. \\

 The value of $\delta_{13}=arg( V^{*PDG}_{ub})$ computed from quark mass ratios and the best values of the parameters $Z^{*1/2}$ and $\Phi^*$ is $\delta^*_{13}=75^{\circ}$ in agreement with the bounds extracted from the precise measurements of the $B^0_d$ oscillation frequency \cite{ref:32} and the measurements of the rates of the exclusive hadronic decays $B^{\pm}\rightarrow \pi K$ and the CP averaged $B^{\pm}\rightarrow \pi^{\pm}\pi^0$ \cite{ref:37}.
It is interesting to notice that, in the flavour symmetry breaking parametrization of the mixing matrix, the best value of the symmetry breaking parameter $Z^{1/2}$ may be written as a purely algebraic number,
\begin{equation}\label{eq:71}
Z^{*1/2}=\frac{1}{2}\left(Z^{1/2}_S- Z^{1/2}_A\right)=\frac{1}{2}\left(\frac{1}{\sqrt 8}+\sqrt 8\right)
\end{equation}
and the best value of the CP violating phase $\Phi$ is consistent with $\Phi^*=\pi/2$.
 
\section*{Acknowledgments}
We are indebted to Dr. M. Mondrag\'on for a careful reading of the manuscript.
This work was partially supported by DGAPA-UNAM under contract No. PAPIIT-IN125298 and by CONACYT (M\'exico) under contract 3909P-E9607.


\begin{thebibliography}{99}

\bibitem{ref:1} A. Mondrag\'on and E. Rodr\'{\i}guez-J\'auregui, {\sl Phys. Rev. }{\bf D59}, 093009, (1999).

\bibitem{ref:2} L.L. Chau and W.-Y. Keung, {\sl Phys. Rev. Lett. }{\bf 53}, 1802, (1984).

\bibitem{ref:3} Particle Data Group, C. Caso {\it et al}. {\sl Eur. Phys.}{\bf J. C 3}, 1, (1998).

\bibitem{ref:4} C. Jarlskog, {\sl Phys. Rev. Lett.}{\bf 55}, 1039 (1985), see also C. Jarlskog,  in ``CP Violation'', edited by C. Jarlskog, Advanced Series on Directions in High Energy Pysics, Vol 3, (World Scientific, Singapore, 1989)

\bibitem{ref:5} H. Fritzsch, {\sl Phys. Lett.}{\bf 70B}, 436, (1977).

\bibitem{ref:6} S. Pakvasa and H. Sugawara, {\sl Phys. Lett.}{\bf 73B}, 61, (1978).

\bibitem{ref:7} H. Fritzsch, {\sl Phys. Lett.}{\bf 73B}, 317, (1978).

\bibitem{ref:8} H. Harari, H. Haut and J. Weyers, {\sl Phys. Lett.}{\bf 78B}, 459, (1978).

\bibitem{ref:9}Y. Chikashige, G. Gelmini, R. P. Peccei and M. Roncadelli, {\sl Phys. Lett.}{\bf B94}, 494, (1980).

\bibitem{ref:10} Y. Yamanaka, H. Sugawara, and S. Pakvasa, {\sl Phys. Rev.}{\bf D25}, 1895, (1982).

\bibitem{ref:11} P. Kaus and S. Meshkov, {\sl Phys. Rev.}{\bf D42}, 1863, (1990).

\bibitem{ref:12} G. C. Branco, J. I. Silva-Marcos and M. N. Rebelo, {\sl Phys. Lett.}{\bf B237}, 446, (1990).

\bibitem{ref:13} H. Fritzsch and J. Plankl, {\sl Phys. Lett.}{\bf B237}, 451, (1990).

\bibitem{ref:14} H. Fritzsch and D. Holtmansp\"{o}tter, {\sl Phys. Lett. }{\bf B338}, 290, (1994).

\bibitem{ref:15} G.C. Branco, and J.I. Silva-Marcos, {\sl Phys. Lett. }{\bf B359}, 166, (1995).

\bibitem{ref:16} H. Fritzsch and Z.Z. Xing, {\sl Phys. Lett. }{\bf B372}, 265,
  (1996).

\bibitem{ref:17} H. Lehmann, C. Newton and T. Wu, {\sl Phys. Lett.}{\bf B384}, 249, (1996).

\bibitem{ref:18} Z.Z. Xing, {\sl J. Phys. }{\bf G23}, 1563,
  (1997).

\bibitem{ref:19} K. Kang and S. K. Kang, {\sl Phys. Rev. }{\bf D56}, 1511,
  (1997).

\bibitem{ref:20} H. Fritzsch and Z. Z. Xing, {\sl Phys. Lett. }{\bf B440}, 313,
  (1998).

\bibitem{ref:21}  J.I. Silva-Marcos, {\sl Phys. Lett. }{\bf B443}, 276, (1998).

\bibitem{ref:22} T. Shinohara, H. Tanaka and I. S. Sogami, {\sl Prog. Theor. Phys. }{\bf 100}, 615, (1998).

\bibitem{ref:23} S. L. Adler, {\sl Phys. Rev. }{\bf D59}, 015012, (1999).

\bibitem{ref:24} R. D. Peccei and K. Wang, {\sl Phys. Rev. }{\bf D53}, 2712, (1996).


\bibitem{ref:25} H. Fusaoka and Y. Koide {\sl Phys. Rev. }{\bf D57}, 3986, (1998).

\bibitem{ref:26} H. Fritzsch, ``Mass Hierarchies Hidden Symmetry and Maximal CP-violation''. hep-ph/9807551.

\bibitem{ref:27} THE BABAR PHYSICS BOOK: PHYSICS AT AN ASYMMETRIC B FACTORY.
By BABAR Collaboration (P.F. Harrison, ed. et al.). SLAC-R-0504, Oct 1998. 1056pp. 
See also the BOOKS subfile under the following call number: QCD201:B3:1998. 
Papers from Workshop on Physics at an Asymmetric B Factory (BaBar Collaboration Meeting), Rome, Italy, 11-14 Nov
1996, Princeton, NJ, 17-20 Mar 1997, Orsay, France, 16-19 Jun 1997 and Pasadena, CA, 22-24 Sep 1997. 

\bibitem{ref:28} H. Leutwyler, {\sl Phys. Lett. }{\bf B378}, 313, (1996).

\bibitem{ref:29} A. Pineda and F.J. Yndurain ``Comment on Calculation of Quarkonium Spectrum and $m_b$, $m_c$ to order $\alpha^4_s$''. hep-ph/9812371.


\bibitem{ref:30} F. Yndur\'ain, {\sl Nucl. Phys. }{\bf B517}, 324, (1998).
 
\bibitem{ref:31} ALEPH Collaboration, Study of tau decays involving kaons, spectral functions and determination of the strange quark mass CERN-EP/99-026, 
hep-ex/9903015.

\bibitem{ref:32} S. Mele, {\sl Phys. Rev. }{\bf D59}, 113011, (1999).

\bibitem{ref:33}A. Ali and D. London,  {\sl Eur.Phys.J.}{\bf C9}, 687-703, (1999).
\bibitem{ref:34}Zoltan Ligeti, Invited talk at Kaon'99, June 21-26, 1999, Chicago
Report-no: FERMILAB-Conf-99/213-T, hep-ph/9908432.

\bibitem{ref:35} Y. Nir, ``Flavor Physics and CP Violation''\\
Lectures given at the 1998 European School of High Energy Physics. University of St. Andrews. Scotland, August 23-september, 1998\\
hep-ph/9810520\\
WIS-98/29/Oct-DPP.

\bibitem{ref:36} A. J. Buras, ``Weak Hamiltonian, CP-violation and Rare Decays'' TUM-HEP-316/98; hep-ph/9806471. \\
To appear in ``Probing the Standard Model of Particle Interactions''\\
F. David and R. Gupta, eds. 1998 Elsevier.


\bibitem{ref:37} M. Neubert, ``Exploring the weak phase $\gamma$ in $B^{\pm}\rightarrow \pi K$ Decays'', hep-ph/9904321.

\end{thebibliography}
\end{document}